# Altitude and Particle Size Measurements of Noctilucent Clouds by RGB Photometry: Radiative Transfer and Correlation Analysis

Oleg S. Ugolnikov

Space Research Institute, Russian Academy of Sciences, Moscow, Russia

E-mail: ougolnikov@gmail.com

**Abstract**

Noctilucent or polar mesospheric clouds have become visually brighter and occurred more frequently during the recent years and decades. The study of possible reasons and relations with climate changes requires data on long-time trends of mean particle size and altitude. Extended worldwide observational data is a good tool for this, and it can be provided by simple RGB-photometry using widely distributed all-sky cameras. Based on observations of bright expanded clouds in summer 2020 and 2021, the method of mean particle size determination is developed, results are validated using the radiative transfer model. The procedure also allows finding the effective "umbral" altitude of clouds. The correlation of size and altitude of particles is compared with existing lidar data and models of particle growth.

**1. Introduction**

Noctilucent clouds (NLC) are optical phenomena observed during high and mid-latitude summers since the end of 19[th] century (Backhouse, 1885; Leslie, 1885). Cloud consists of small ice particles in the summer mesosphere at 80-85 km altitudes; temperature below 150K is necessary to form the ice in this layer. Initially noted as a rare event, NLC were observed more frequently in the 20[th] century (Thomas and Olivero, 2001). The possible reason was mesosphere cooling by radiative effects caused by greenhouse gases, especially $CO_2$ (Roble, Dickinson, 1989). In this case, NLC can be an indicator of global climate change.

However, visual occurrence analysis is not exact for many reasons. Moreover, no visual trends were observed in the late 20[th] – early 21[st] century (Pertsev et al., 2014). Direct relation with temperature changes is also doubtful since the temperature is now practically constant in the upper summer mesosphere. Figure 1 shows the profiles of mean June-July temperature and its trend by 20 years of measurements by SABER instrument onboard TIMED satellite (Russell et al., 1999), providing the best vertical resolution in the mesosphere. Data is averaged on the locations with latitudes ±3° and longitudes ±15° from the observation point of this paper (55.6°N, 36.6°E). We see that the upper mesosphere is practically the only layer above the tropopause where negative temperature trend vanishes; this is related to the compensation of radiative cooling by absorption of upcoming emission (Berger and Lübken, 2011).

However, negative trend appears at 80 km and below; this can be interpreted as a downshift of the upper mesosphere related to atmosphere shrinking caused by cooling in a wide range of altitudes (Berger and Lübken, 2015). A higher $H_2O$ mixing ratio characterizing lower layers is the principal for the growth of large ice particles with a radius of 50 nm and more (Lübken et al., 2018). Due to the strong dependence of scattering properties of particles on radius $a$ ($\sim a^6$), large particles are the principal for creating bright visual clouds. However, we should note that large particles are not formed in the layers where the temperature is significantly less than ice frost temperature, appearing near the lower border of the frost layer (Merkel et al., 2009).

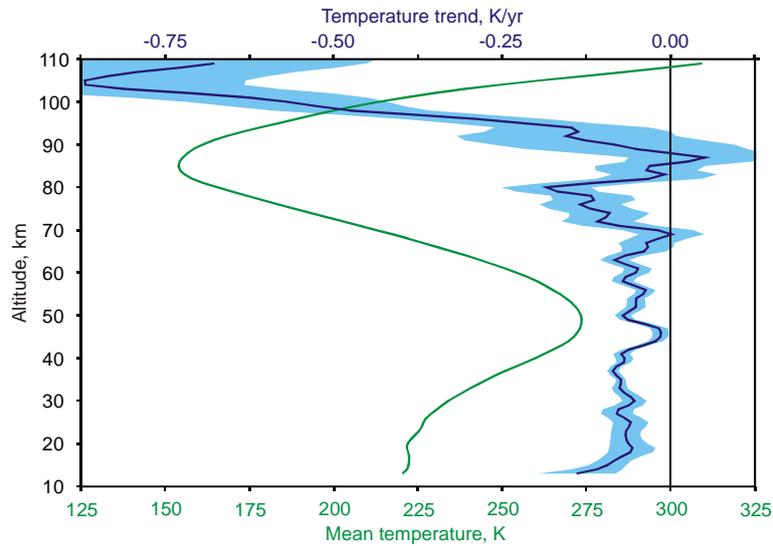

*Figure 1. Mean June-July temperature and its trend by TIMED/SABER measurements in 2002-2021 in the vicinity of the observation site.*

If climate changes in the atmosphere drive the increase of NLC occurrence rate, then we will fix the trends of observable characteristics of clouds: increase of mean particle size and decrease of mean altitude, this is confirmed by the model results (Berger and Lübken, 2015; Lübken et al., 2018). To extend the measurement data, we should use the simple and cost-effective technique of long-term altitude and particle size measurements that can be potentially implemented in a large number of locations on Earth. This technique can be photometry of expanded cloud fields by wide-angle RGB cameras. The size determination method in its initial form was suggested by Ugolnikov et al. (2017) and applied to a single case of bright NLC at 68°N. Here we test the basic assumptions of the procedure by radiative transfer analysis and find the altitude among with particle size.

We apply all procedures described above to the observations in 2020 and 2021. These summers were characterized by many bright NLC occurrences in mid-latitude Russia, indicating the relevance of trend study again. Results on size and altitude measurements during these events allow studying the relationship between these values and comparing it with other measurements and theoretical models.

**2. Observations**

*2.1. Overview*

RGB measurements of sky background were held in central Russia (55.6°N, 36.6°E) by two identical Xiaomi Mi Sphere Cameras. One lens of the camera has a field of view about 190°, capturing the whole sky hemisphere as the axis was directed close to the zenith. The exact camera position and field parameters are defined from stars images in the night frames. Star photometry is also used to find the atmosphere transparency; see below. Cameras worked in raw output format from sunset to sunrise, expose times in the twilight period when NLC are observed varied from 0.25 to 4 seconds and increased to 8 seconds during the night. The data of the sky part with zenith angles up to 65° was processed. The angular resolution of the camera close to the zenith is 19 pixels per degree, very slightly decreasing to 18 pixels per degree by zenith angle at the edge of the processed sky area.

This work is based on six occurrence events of bright NLC: the evening of June 21 and morning of July 6 (7 by local time), 2020; the morning of June 23 (24 by local time), evenings of June 29 and July 4, and morning of July 9 (10 by local time), 2021. It is worthy of note that four 2021 events along with one previous (morning, June 18 by local time, missed in the analysis due to intermixing cirrus clouds in the sky) are separated by 5-6 days one from another, showing the 5-day planetary wave passage (Gadsden, 1985; Merkel et al., 2003; Stevens et al., 2017).

The clouds were expanded far beyond the zenith in each case, covering a wide range of scattering angles; this is important for the method described in this work. In several cases, for example, in the evening of June 29, 2021, bright NLC completely covered the sky until immersion into the umbra of the Earth (see Figure 2). Both cameras recorded all six events; for better analysis of method accuracy, the mean size and altitude of NLC are determined separately based on the data of each camera.

*2.2. RGB spectral bands*

Blue, green, and red spectral bands of color cameras are wide and must be correctly considered during data procession. Note that NLC particles are significantly less than optical wavelengths, and scattering coefficient $S$ dependency is close to Rayleigh law: $S \sim \lambda^{-4}$. Effective wavelengths $\lambda_{RGB}$ for NLC observations are shorter than effective wavelengths of RGB detectors themselves. We define spectral characteristics of bands 1(B), 2(G), and 3(R) for NLC observations as

$$F_{1,2,3}(\lambda) = SOL(\lambda) \cdot F_{D1,2,3}(\lambda) \cdot \lambda^{-4} \cdot e^{-\tau(\lambda)/\cos Z_0} \quad (1).$$

Here $SOL$ is the solar energetic spectrum, $F_{Di}$ is energetic sensitivity of detector at the wavelength $\lambda$, $\tau$ is atmospheric vertical optical depth (sum of Rayleigh, ozone and aerosol optical depths), $Z_0$ is the mean zenith angle of observation point in the sky, we take it to be equal to 45°. Profiles $F_{1,2,3}(\lambda)$ are shown in Figure 3.

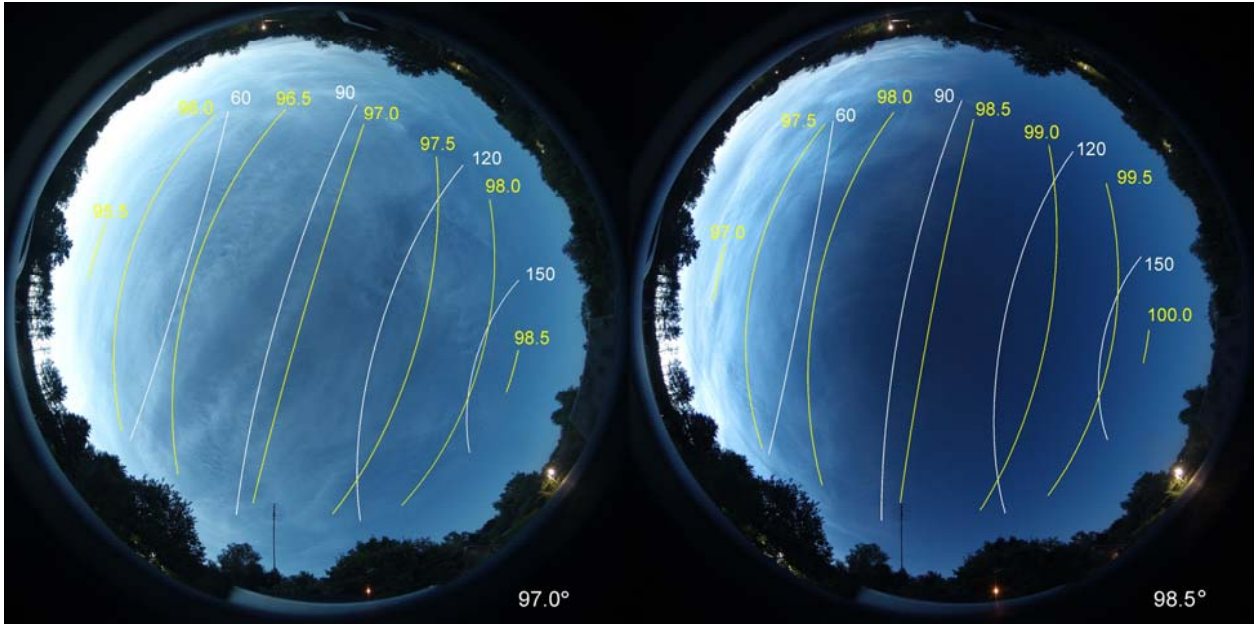

*Figure 2. RGB images of the sky with noctilucent clouds in the evening of June 29, 2021, at solar zenith angle 97.0° and 98.5°. The lines of equal scattering angle are shown in white, the lines of equal local solar zenith angle are shown in yellow.*

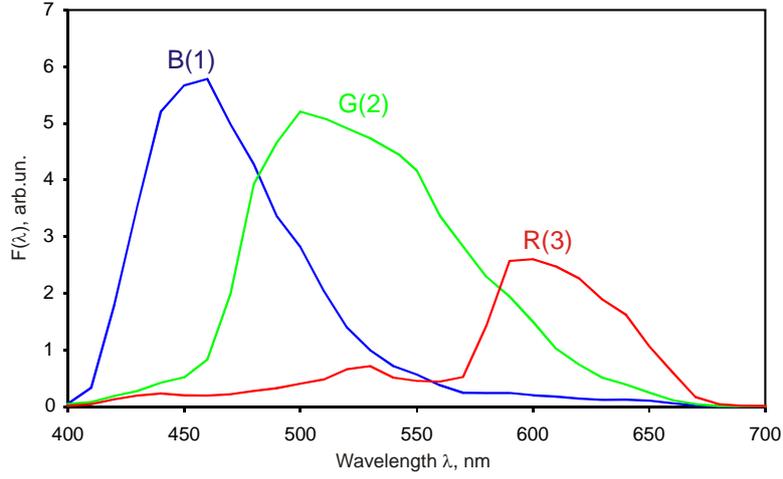

*Figure 3. Instrumental RGB bands with an account of the solar spectrum, atmospheric transparency, and scattering properties of small particles.*

The traditional definition of effective wavelength as an average value with weights proportional to $F(\lambda)$ is not optimal for this work. Since the observable effect in NLC used to find the mean particle size is proportional to $\lambda^{-2}$ (Ugolnikov et al., 2017), we define "effective back-squared" wavelength as follows:

$$\frac{1}{\lambda_{1,2,3}^2} = \frac{\int \frac{1}{\lambda^2} F_{1,2,3}(\lambda) d\lambda}{\int F_{1,2,3}(\lambda) d\lambda} \qquad (2).$$

These values are equal to 469, 523, and 578 nm for blue, green, and red bands, 5, 6, and 9 nm less than the "traditional" effective wavelength in these bands, respectively. Rayleigh extinction and ozone Chappuis absorption significantly change inside each of three bands; we do not use the effective wavelength terms for accurate analysis in this paper; it is proceeded by integration by the profile $F(\lambda)$.

*2.3. NLC separation on the twilight background*

The signal-to-background separation procedure for NLC is a pretty complicated problem. NLC have no sharp borders and can cover the sky hemisphere completely. The procedure is based on the variability of NLC brightness in the sky. The first and most effortless approach was the Fourier analysis of sky brightness along definite lines in the sky, excluding low-order components (Ugolnikov et al., 2017). However, this procedure leads to significant signal loss while the sky background is not deleted completely. A more efficient technique was suggested by Ugolnikov et al. (2021) to analyze polar stratospheric clouds. We take profiles of sky brightness along the line of constant scattering angle; examples of such lines are shown in white in Figure 2. Having binned the data in 1°-boxes, we divide the profile into the points associated with NLC and smoothed background. Mathematical details of the procedure are described in the paper referred above.

Neighbor NLC points of one profile are united into "spots". For each spot, we define the NLC brightness in three bands, $I_{1,2,3}$; the ratio of NLC and background brightness, $\eta_{1,2,3}$; spot mean zenith angle $Z$, scattering angle (constant along the profile), $\theta$, and local solar zenith angle $z_L$. To find the last value, we must know the approximate altitude of NLC (82 km). The lines of constant $z_L$ are shown in yellow in Figure 2; for NLC in the zenith, this value is equal to solar zenith angle $z$ in the observation site. The spots with angular length more than 8° and relative NLC brightness $\eta_{1,2,3} > 0.02$ are included in the following analysis.

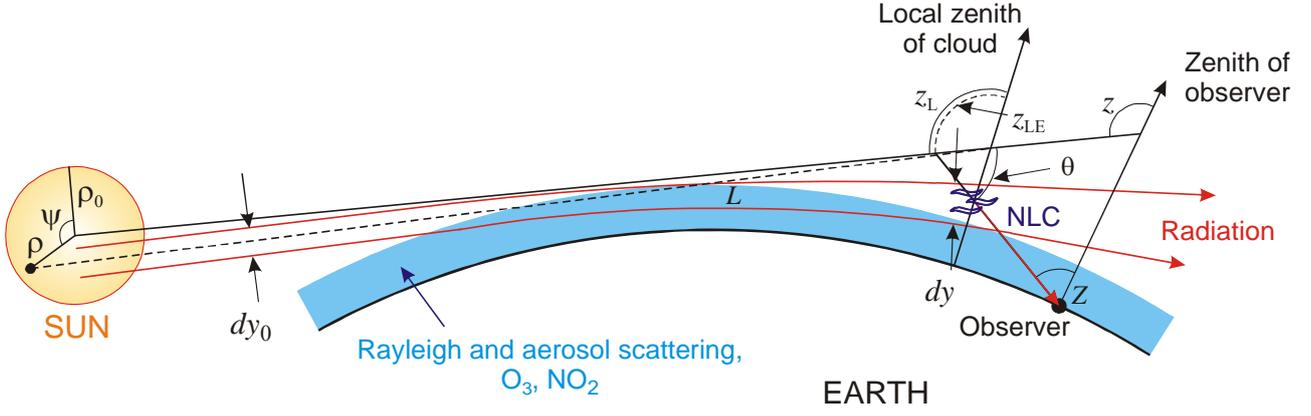

*Figure 4. Scheme of transfer and scattering of solar emission by NLC.*

## 3. NLC color properties

Figure 4 shows the optical scheme of NLC illumination. The primary observational parameters of cloud element are its color ratios $I_{2,3}/I_1$, not the brightness itself. The factors influencing NLC color are following:
   a) Solar radiation transfer through the atmosphere before scattering. The factor is defined by the value $z_L$ – solar zenith angle observed from the cloud;
   b) Color properties of scattering. The factor is defined by wavelength, particle size, and scattering angle θ. Since the refraction of solar emission is less than 1°, we can assume that the scattering angle is equal to the angular distance between NLC spot and the center of the Sun in the sky (see Figure 4).
   c) Extinction of scattered light in the lower atmosphere. Far from the horizon, the factor is defined by Bouger law and depends on the difference of vertical optical depths τ in bands 1 and 2(3) and zenith angle of cloud element, $Z$.

Ugolnikov et al. (2017) suggested the linearization scheme to separate all factors and find the particle radius $a$. It uses the fact that the ratio of scattering functions of small particles in different color bands (2 or 3 and 1) simply depends on scattering angle:

$$\frac{S_{2,3}(\theta, a)}{S_1(\theta, a)} = \frac{S_{2,3}(\pi/2, a)}{S_1(\pi/2, a)} \cdot (1 + P_{2,3}(a)\cos\theta) \tag{3}$$

Here coefficient $P$ is defined by the particle size. Its dependence on radius $a$ for monodisperse particles is shown in Figure 5. For even smaller particles, the analytical expression can be written:

$$P_{2,3} = \frac{m+1}{5} \cdot \left( \frac{4\pi^2 a^2}{\lambda_{2,3}^2} - \frac{4\pi^2 a^2}{\lambda_1^2} \right) \tag{4}$$

Here $m$ is the refractive index of ice; it equals 1.31 for NLC case (Iwabuchi, Yang, 2011). This dependency is also shown in Figure 5 by dashed lines. We see that expression (4) is valid for the particle radius up to 60-70 nm. However, larger particles are possible in NLC, and we use the general expression (3) calculating coefficients $P$ by Mie theory with integration by profiles $F(\lambda)$.

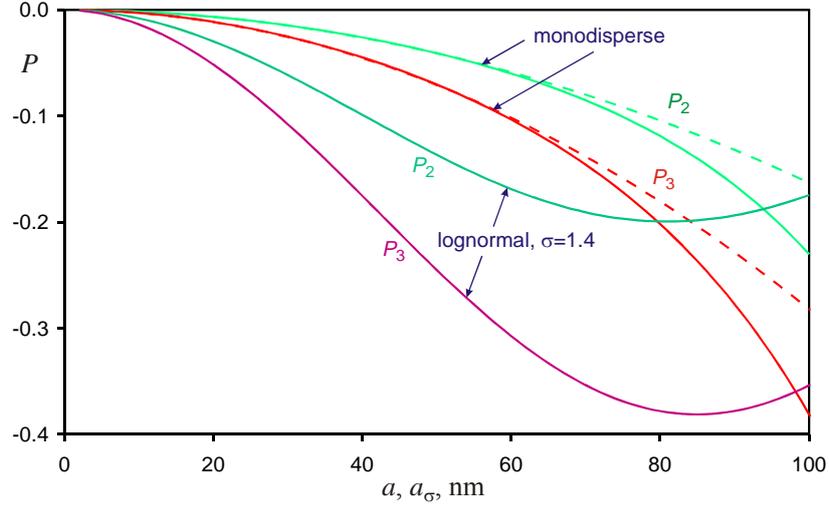

*Figure 5. The color gradient of ice particles scattering functions for monodisperse and lognormal particle distributions. Dashed lines correspond to approximation (4).*

Taking into account all three factors defining NLC color listed above, we can write the expression for the ratio of NLC fragment intensity in two bands:

$$\frac{I_{2,3}}{I_1}(z_L, \theta, Z) = K_{2,3}(z_L) \cdot \frac{S_{2,3}(\pi/2)}{S_1(\pi/2)}(1 + P_{2,3} \cos\theta) \cdot e^{-\frac{\tau_{2,3} - \tau_1}{\cos Z}}$$

(5).

Here, factor $K$ represents the change of the color ratio of solar radiation while propagating through the atmosphere to NLC; the last term is related to the extinction of NLC radiation in the lower atmosphere above the observer. Using the slight change of NLC color by each of these factors, Ugolnikov et al. (2017) suggested the direct linearization of this equation. A more exact approach is a linear logarithmic expression:

$$\ln\frac{I_{2,3}}{I_1}(z_L, \theta, Z) = \ln\frac{I_{2,3}}{I_1}(z_{L0}, \pi/2, Z_0) + Q_{2,3}(z_L - z_{L0}) + P_{2,3}\cos\theta + T_{2,3}\left(\frac{1}{\cos Z} - \frac{1}{\cos Z_0}\right);$$

$$Q_{2,3}(z_L - z_{L0}) = \ln\frac{K_{2,3}(z_L)}{K_{2,3}(z_{L0})}; \qquad T_{2,3} = \tau_1 - \tau_{2,3}$$

(6).

Here $Z_0=45°$, $z_{L0} = 96°$. Ugolnikov et al. (2017) took $z_{L0} = 97°$, the reason to change this reference point is discussed below, and used the oppositely defined $T$ values. The only approximation made here is the linearization of the term $\ln(1+P\cos\theta)$, which can be done since the module of $P$ rarely exceeds 0.1 (see Figure 5). Summarizing this, we can list the basic assumptions of this simplified scheme of finding the value of $P$ and then the particle size:

a) Low contribution of multiple scattering or scattering of dusk segment emission by NLC;
b) Linear dependency of color ratio logarithm of Sun illumination of NLC ($K$) by solar zenith angle $z_L$.

The assumption (a) is confirmed by polarization measurements of NLC (Ugolnikov et al., 2016; Ugolnikov and Maslov, 2019) and distinct edge of Earth's umbra in NLC field (see Figure 2, right). The most complicated question is the assumption (b). It can be tested by radiative transfer model with an account of refraction, the angular size of the Sun and its brightness distribution, Rayleigh and aerosol scattering, absorption by stratospheric $O_3$ and $NO_2$ in the optical spectral range. This model will be considered in the following chapter of the paper.

## 4. Radiative transfer model

Let the fragment of NLC is seen at zenith angle $Z$ (see Figure 4, the case of NLC in the solar vertical is shown for simplicity). It is illuminated by the fragment of the solar disk with coordinates $\rho$ and $\psi$. The local zenith angle of the center of the Sun visible from the cloud is $z_L$, local zenith angle of the considered fragment of the Sun is $z_{LE}(z_L,\rho,\psi)$. Wavelength $\lambda$, the value of $z_{LE}$, and cloud altitude $H$ completely define the trajectory $L$ and radiative transfer from the solar surface element to the cloud. Energy flux incoming to the atmosphere from this element from the Sun is

$$j_0(\lambda,\rho) = SOL(\lambda) \cdot s(\lambda,\rho); \quad \int_0^{\rho_0} 2\pi\rho \cdot s(\lambda,\rho)\, d\rho = 1 \tag{7}$$

Here $s(\lambda,\rho)$ is the normalized distribution of brightness on the solar disk; data is taken from (Allen, 1973), $\rho_0$ is the angular radius of the Sun. Passing through the atmosphere to the NLC, the flux weakens:

$$j(\lambda,\rho,\psi) = j_0(\lambda,\rho) \cdot \frac{dy_0}{dy} \cdot exp\left(-\int_L E(\lambda,h)\,dl\right) = j_0(\lambda,\rho) \cdot k(\lambda, z_{LE}(z_L,\rho,\psi), H) \tag{8}$$

Here the integral is total optical depth by the trajectory $L$, $E$ is extinction coefficient at the altitude $h$ of the trajectory element $dl$. Since the solar radiation effective for NLC illumination is transferred mainly through the stratosphere, extinction $E$ is defined basically by Rayleigh and stratospheric scattering and Chappuis ozone and nitric dioxide absorption. We can run the integration procedure and calculate $k$ values using MLS/Aura satellite data on temperature and ozone and Suomi NPP/OMPS limb profile data for the stratospheric aerosol for the date of NLC observation and location close to the solar emission perigee (NASA GES DISC, 2021). For $NO_2$, which influence to NLC color is less, we use typical altitude profiles for NLC season obtained in the location close to the observation site (Gruzdev and Elokhov, 2021) with an account of their dependence on the solar zenith angle (calculated for each point along the trajectory) and difference of evening and morning twilight conditions.

Calculations are performed for different values of $\lambda$, $z_{LE}$, and $H$ with an account of refraction as a function of wavelength (Allen, 1973) and altitude, the term $dy_0/dy$ in equation (7) represents the divergence of refracted emission, see Figure 4. For these calculations, we use the model of spherical Earth with the radius equal to the curvature radius of Earth ellipsoid in the plane of solar vertical during NLC observations at given latitude and time (6385 km for $z=97°$).

We integrate the energy flux by the solar disk and spectral profiles $F_i(\lambda)$. Taking into account equation (1), we find the model color ratio of NLC for $\theta=\pi/2$ and $Z=Z_0$:

$$\frac{I_{M2,3}}{I_{M1}}(z_L,\pi/2,Z_0) = \frac{\int_\lambda F_{2,3}(\lambda)\int_0^{\rho_0} s(\lambda,\rho)\int_0^{2\pi} \rho \cdot k(\lambda,z_{LE}(z_L,\rho,\psi),H)\,d\psi\,d\rho\,d\lambda}{\int_\lambda F_1(\lambda)\int_0^{\rho_0} s(\lambda,\rho)\int_0^{2\pi} \rho \cdot k(\lambda,z_{LE}(z_L,\rho,\psi),H)\,d\psi\,d\rho\,d\lambda} \equiv K_{M2,3}(z_L,H) \tag{9}$$

Function $K_M(z_L)$ is a model analog of $K(z_L)$ in equation (5) differing from it by a constant factor. Figure 6 shows dependencies $K_M(z_L)$ for clouds altitude $H$ equal to 80, 82, and 84 km (blue lines, dependency for 82 km is marked in bold). We can see three basic stages of NLC visibility during the twilight:

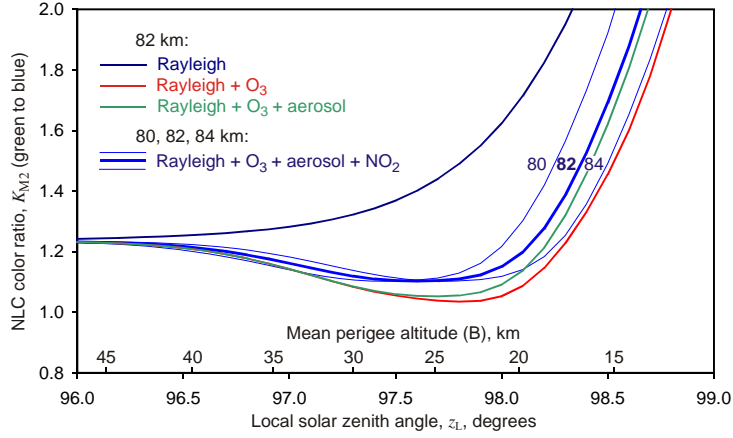

*Figure 6. Color ratio $K_{M2}$ of noctilucent clouds for $\theta=\pi/2$ and $Z=Z_0$ by model calculations for the evening of June 29, 2021.*

1) While the local solar zenith angle is less than 96.0°-96.2°, radiation of Sun reaches NLC through upper atmospheric layers being practically unabsorbed, the color of NLC remains constant. It is the reason to use $z_{L0} = 96°$ in the linearized approach (6).
2) When Sun depresses below to the zenith angle up to 97.5°, solar radiation trajectory shifts down to the stratospheric ozone layer, and NLC get slightly bluer owing to the spectral properties of $O_3$ Chappuis band. This effect is weakened by Rayleigh scattering and $NO_2$ absorption in the blue part of the spectrum.
3) Finally, during the dark twilight NLC are emitted by the Sun through lower atmospheric layers and get redder due to Rayleigh scattering; aerosol also contributes to this.

The influence of each factor on NLC color is seen in Figure 6, where the dependency $K_{M2}(z_L)$ is also shown for the atmosphere with Rayleigh scattering only, Rayleigh scattering with ozone, and Rayleigh and aerosol scattering with ozone, $H$=82 km. The altitude of the mean perigee of the ray in 1(B) channel is also shown on the *x*-axis. Color evolution of NLC is also seen in Figure 2 (right).

We should note that while Rayleigh scattering does not dominate during the first two periods, the dependency $K_M(z_L)$ is relatively weak and can be considered linear; this leads to errors in a color ratio not more than 5%. This means that the linearization approach (6) can be used for NLC size determination if the color data is restricted by solar zenith angle: $z_L$<97.5°. This threshold value is close to the $K_M(z_L)$ minimum for cloud altitude $H$=82 km.

However, this can be further clarified if the mean NLC altitude is known and *K* value is calculated from the model. We can find NLC umbral altitude comparing model and observation dependencies of color ratio on $z_L$, using the fact that they strongly depend on *H* during the dark twilight period, $z_L$>97.5°. Simultaneous altitude and particle size measurements can be helpful to check models of NLC particles nucleation.

## 5. NLC umbral altitudes

Figure 7 shows the measured color ratios $I_{2,3}/I_1$ at June 29, 2021, by one camera; the data with cloud to background ratio $\eta_1$>0.05 is selected. We see that the color evolution agrees with model dependency $K_M(z_L)$. Color changes slowly until solar zenith angle reaches 97.5° and turns redder after that owing to Rayleigh and aerosol scattering of solar radiation in the lower atmosphere. This process significantly depends on NLC altitude and thus can be the basis for altitude determination.

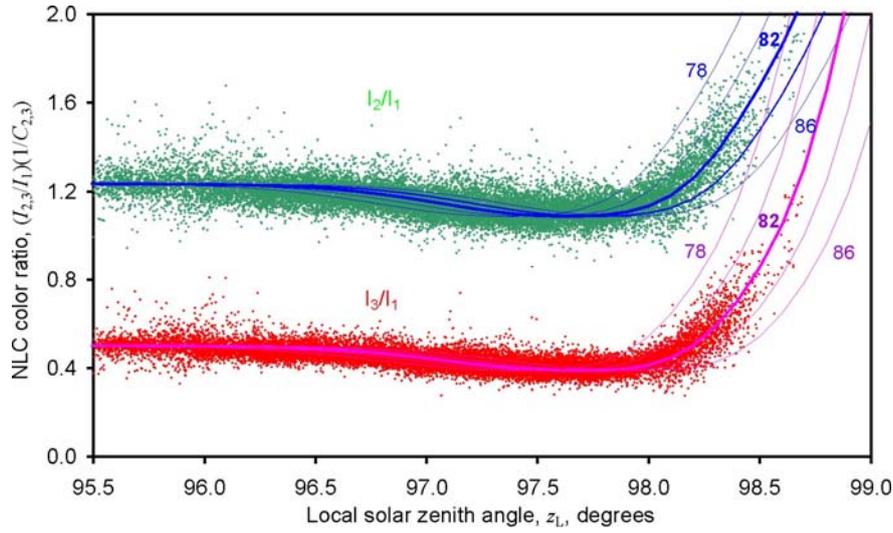

*Figure 7. Measured color ratios of NLC (evening, June 29, 2021) compared with model dependencies for different NLC altitudes.*

However, this must be done not by directly comparing observational and theoretical data in Figure 7. Experimental data on NLC color are obtained at different point zenith angles $Z$ and scattering angles $\theta$. Effect of scattering functions and lower atmospheric extinction cause the small but noticeable change of color ratios. We use the linearization scheme similar to (6) to take these effects into account:

$$ln\frac{I_{2,3}}{I_1}(z_L,\theta,Z) = ln\,C_{2,3} + ln\,K_{M2,3}(z_L,H) + \overline{P}_{2,3}\cos\theta + \overline{T}_{2,3}\left(\frac{1}{\cos Z} - \frac{1}{\cos Z_0}\right) \quad (10).$$

Here free parameter $C$ differs a little from unity owing to variations of atmospheric transparency at solar zenith angle $Z_0$ and deviations of wavelength dependency of scattering coefficient $S(\pi/2)$ from $\sim\lambda^{-4}$ if particles are quite large, see equation (5); the data in Figure 7 is corrected by this factor. Applying this to each NLC spot and taking the weighting system proportional to the measured cloud to background brightness ratio $\eta_1$, we find the umbral NLC altitude, $H$ by the least-squares method.

### 6. Validation of particle size estimation procedure

The particle radius definition described above is based on the property of scattering functions (3) and search for parameter $P$ by linearized least-squares method (6). Accuracy of this approach and possible systematic errors is the question of a particular study. Another problem is that for any definite twilight stage, the scattering angle $\theta$ and local solar zenith angle $z_L$ change almost mutually in the sky (see Figure 2); the same can be true for spot zenith angle $Z$ in the dusk segment where NLC are brightest. This problem can cause the increase of errors of the linearized procedure. It is solved by extension of observation period during the twilight; however, it can still affect the final result.

Analysis of Rayleigh and aerosol scattering and $O_3$ and $NO_2$ absorption of solar radiation made in this paper had shown that particle size determination procedure could be based on measured NLC colors during the light twilight stage $z_L<97.5°$ when color is changing slowly and less dependent on the variations of tropospheric and stratospheric aerosol.

Testing this procedure, we run it for $z_L<97.5°$ in two forms. The first one is general and defined by equation (6). For the second one, we use independent data on other factors influencing NLC color: radiation extinction before and after scattering on NLC particles. The first factor ($K_M$) is calculated by the model described by equations (7-9) based on altitude of NLC $H$ found above, MLS Aura data on temperature and ozone, OMPS data on aerosol, and $NO_2$ profiles by (Gruzdev and Elokhov, 2021).

Local atmospheric transparency (values of $T_S$) can be estimated by stellar photometry on nighttime images. Since we need to obtain the data in a wide range of zenith angles during the short summer night, we use a number of stars with identical spectra close to the spectrum of NLC. We had chosen eight stars of early spectral classes B9-A0-A1 with magnitudes brighter than $3^m$, widely distributed in the sky. The brightest star is Vega (α Lyr, A0, $0.03^m$). Finally, we can find NLC color ratios corrected for light extinction and absorption:

$$\left(\frac{I_{2,3}}{I_1}\right)_C(\theta) = \frac{I_{2,3}}{I_1}(z_L,\theta,Z) \cdot \frac{1}{K_{M2,3}(z_L,H)} \cdot exp\left(-T_{S2,3} \cdot \left(\frac{1}{\cos Z} - \frac{1}{\cos Z_0}\right)\right) \quad (11).$$

These colors should change only due to variation of scattering angle θ. Figure 8 shows the example of such dependency for NLC event on June 29, 2021, for NLC spots with the sky to background ratio $\eta_1>0.1$. A gradient of a color ratio by $\cos\theta$ is negative, showing that $P<0$. It can be seen from approximated equation (4) for this value, $\lambda_{2,3}>\lambda_1$. Now we can find the refined values $P_R$ by the simple two-parametric least square method:

$$\left(\frac{I_{2,3}}{I_1}\right)_C(\theta) = \left(\frac{I_{2,3}}{I_1}\right)_C(\pi/2) \cdot (1 + P_{R2,3}\cos\theta) \quad (12).$$

Knowing $P$ obtained from linear analysis (6) or $P_R$ just found, we can find the effective radius of particle $a$ ($a_R$). We use Mie theory data shown in Figure 5, not the approximate expression (4). We can also find the median radius of particle for lognormal distribution $a_\sigma$ ($a_{\sigma R}$) with width σ=1.4, typical for NLC (von Savigny and Burrows, 2007); the corresponding dependency is also shown in Figure 5. If this radius is not higher than 50 nm, the scattering properties will be close to the monodispersed ensemble with $a = 2a_\sigma$.

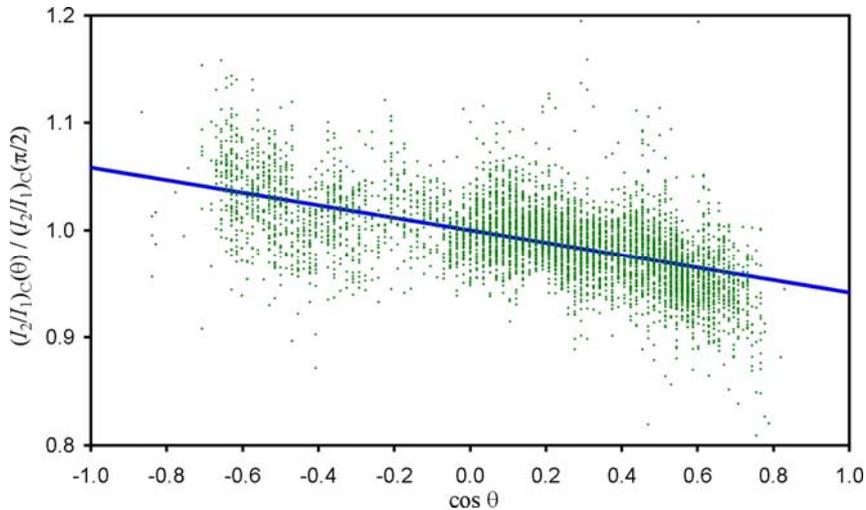

*Figure 8. Dependence of corrected NLC color (2 to 1) on $\cos\theta$, the evening of June 29, 2021.*

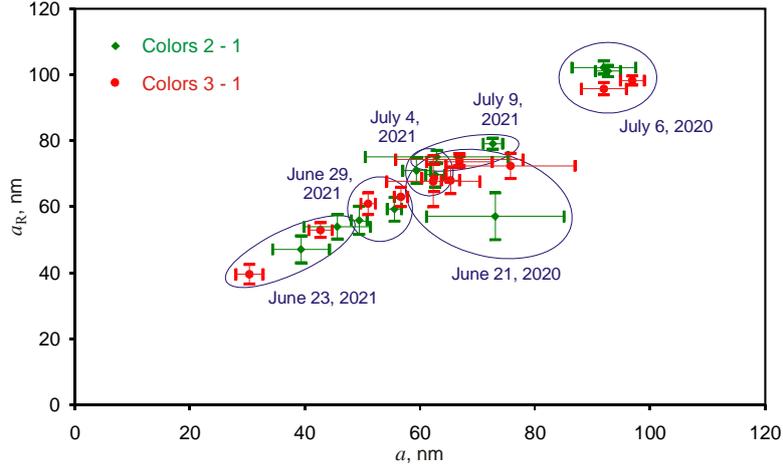

*Figure 9. Comparison of effective particle radius (monodisperse ensemble, a) estimated by linearized procedure (6) and refined method ($a_R$). Data of the same twilight is united in blue ellipses.*

## 7. Results

A comparison diagram of effective particle radii *a* obtained by general procedure (6) and $a_R$ by refined method (12) is shown in Figure 9. We have four estimations obtained by two cameras in two different color pairs (2-1 and 3-1) for each NLC occurrence. They are shown separately, being united in blue ellipses. An error of refined radius $a_R$ is also contributed by errors of stellar photometry while finding parameter $T_S$. We can conclude that the general method can be used for the size definition of noctilucent clouds, although it gives slightly lower particle radius values than the refined method.

Figure 10 shows the relation of the refined effective radius of particles $a_R$ and umbral altitude of NLC found by least-squares analysis of equation (10), *H*. Scale of median particle radius for lognormal distribution $a_{\sigma R}$ is also shown. The example of the model mean radius profile of Leibniz Institute Middle Atmosphere Model (LIMA, Lübken et al., 2018) and average lidar mean particle size profile (Baumgarten et al., 2010) are shown for comparison. We see the general properties of "size-altitude" dependency: particle size reaches the maximum near the lower border of the frost layer while its altitude can change. Remarkably, this altitude is lower while particles are larger in observational (lidar and RGB photometry) cases. The downshift of NLC to the layers with higher water vapor concentration is the primary factor leading to the appearance of bright clouds with large particles. The same is confirmed by NLC polarimetry (Ugolnikov and Maslov, 2019), where the clouds with an effective radius above 100 nm appeared at the bottom of the frost layer near 80 km.

Another critical question is the sensitivity of results to the basic atmosphere characteristics fixed in the model. Table 1 contains the shift values of NLC umbral altitude and effective particle radius due to a 10% increase of Rayleigh scattering, aerosol scattering, $O_3$, and $NO_2$ concentration at all altitudes in the atmosphere. It is no surprise that Rayleigh scattering and ozone absorption are the primary factors defining the radiative transfer and, thus, the measured value of NLC umbral altitude. Sensitivity to aerosol changes is not strong; this is explained by relatively flat (~$\lambda^{-2}$ on average) wavelength dependency of stratospheric aerosol extinction according to OMPS data. Sensitivity to stratospheric $NO_2$ variations is about one-half of sensitivity to aerosol, and this atmospheric component must be taken into account in the color analysis of NLC. The last row in the table shows the altitude shift if the Sun is considered a point light source. This shift is as high as 0.5 km, showing the necessity of solar disk integration with an account of darkening to the edge.

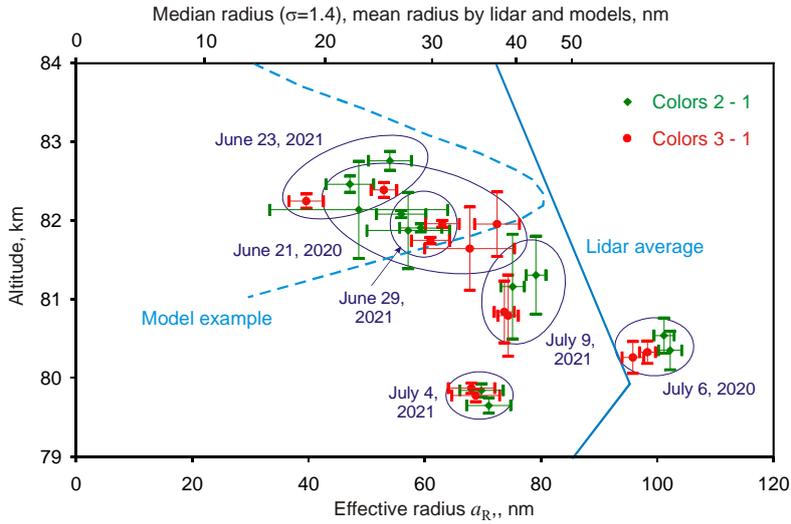

*Figure 10. Diagram "altitude – particle size" for RGB observations compared with lidar measurements (Baumgarten et al., 2010) and model example (Lübken et al., 2018).*

While the umbral altitude of NLC measured by RGB photometry is quite sensitive to atmospheric characteristics, the mean particle size is not, as we see in the right part of the table. The reason is the almost constant color ratios $K(z_L)$ of NLC while local solar zenith angle $z_L$ is less than 97.5°. This fact is also the reason for good general analysis results (6), requiring the linearity of this dependence. It is the main outcome showing the possibility of systematic measurements of mean particle size of NLC by wide-angle color cameras.

## 8. Discussion and conclusion

In this paper, we had considered the technique of noctilucent cloud particle analysis by photometry by RGB all-sky camera; this is the simplest and less expensive method of NLC study. Analysis of six bright NLC occurrence events in 2020 and 2021 had shown that the effective size of particles could be found by a simple linearized approach based only on observational data. Here these results were confirmed by deeper analysis with the integral model of radiation transfer to NLC and stellar measurements of atmospheric extinction above the observer.

Another goal of the paper is to find the mean "umbral" altitude of NLC based on its color changes as clouds immerse into the shadow of the ozone layer and then lower atmosphere with intensive Rayleigh scattering. For detectors with wide spectral bands like RGB cameras, the effect of ozone is significant, but it can be even stronger if we choose a narrow band near the maximum of Chappuis absorption (about 600 nm). In this case, the ozone content in the stratosphere can be considered a free parameter and measured alongside particle size and altitude of NLC.

| Variation | $\Delta H$ (2-1), km | $\Delta H$ (3-1), km | $\Delta a$ (2-1), nm | $\Delta a$ (3-1), nm |
|---|---|---|---|---|
| Rayleigh, +10% | +0.55 | +0.53 | −1.0 | −1.0 |
| $O_3$, +10% | −0.63 | −0.57 | +2.2 | +2.1 |
| Aerosol, +10% | +0.11 | +0.13 | −0.1 | −0.1 |
| $NO_2$, +10% | +0.07 | +0.06 | −0.3 | −0.3 |
| Sun is a point source | +0.48 | +0.42 | −0.3 | −0.2 |

*Table 1. The shift of retrieved values of umbral altitude and effective particle size of NLC resulted from variation of model parameters of the atmosphere and radiative transfer (example for June 29, 2021).*

Diagram "median particle radius – altitude" confirms lidar results with a maximum size near 80-81 km. In the case of negative temperature trends below NLC layer, this altitude is expected to decrease while maximum particle size can increase. The simple technique described in this paper can be the basis of multi-point or net observations of NLC by all-sky cameras widely distributed in recent years. Measurements by cameras at a distance of 50-300 km one from another will also make it possible to determine the altitude of NLC by well-known triangulation techniques.

**Acknowledgments**

The author is grateful to Sergey V. Pilipenko (Astro-Space Center of Physical Institute of Russian Academy of Sciences) and Igor A. Maslov (Space Research Institute, Russian Academy of Sciences) for help in the organization of measurements. The author also thanks Oleg V. Postylakov and Alexander N. Gruzdev (Institute of Atmospheric Physics, Russian Academy of Sciences) for help in the analysis of $NO_2$ in the stratosphere.